\documentstyle[psfig,aps,eqsecnum]{revtex}
\newcommand{\beqa}{\begin{eqnarray}}
\newcommand{\eeqa}{\end{eqnarray}}
\newcommand{\non}{\nonumber}

\newcommand{\um}{\frac{1}{2}}
\newcommand{\mj}{m\sqrt{\frac{2}{j(j+1)}}}
\begin{document}

\title{Structure of the vacuum states in the presence of 
isovector and isoscalar pairing correlations}

\author{D. R. Bes $^{\rm 1,2,4}$, 
O. Civitarese $^{\rm 1,3}$, 
E. E. Maqueda $^{\rm 1,2}$ and 
N. N. Scoccola $^{\rm 1,2,4}$}

\address{$^{\rm 1}$ Consejo Nacional de Investigaciones 
Cient{\'\i}ficas y T\'ecnicas, Argentina}

\address{$^{\rm 2}$ Departamento de F\'{\i}sica, 
CNEA, Avda. Gral. Paz 1499, (1650) San Mart{\'\i}n, Argentina}

\address{$^{\rm 3}$ Departamento de F\'{\i}sica, 
UNLP, C. C. 67, (1900) La Plata, Argentina}

\address{$^{\rm 4}$ Universidad Favaloro, Sol{\'\i}s 453, 
(1078) Buenos Aires, Argentina}

\date{\today}
\maketitle

\begin{abstract}

The long standing problem of proton-neutron pairing and, in particular,
the limitations imposed on the solutions by the available symmetries,
is revisited.  We look for solutions with non-vanishing expectation
values of the proton $\Delta_p$, the neutron $\Delta_n$ and the
isoscalar $\Delta_0$ gaps. For an equal number of protons and neutrons
we find two solutions with $\Delta_p=\pm \Delta_n$, respectively. The
behavior and structure of these solutions differ for spin saturated
(single $l$-shell) and spin unsaturared systems (single $j$-shell). In
the former case the BCS results are checked against an exact
calculation.

\end{abstract}

\section{Introduction} \label{intro}

There is presently a revival of the interest on the solutions to the
old problem of proton-neutron pairing. This is due to the  availability
of unstable beams and targets and of large array detectors, which allow
the experimental study of heavier nuclei with equal number of protons and
neutrons. Most of the recent papers include rather complete histories
of the subject (see, for instance, refs. \cite{Go99,Sp97}).

Specific aspects of the formalism may be found in a review article by
Goodman\cite{go72}. In particular, the topic of the limitations on the
form of the solutions which are imposed by the available symmetries, is
covered there. Since then, most papers on the neutron-proton pairing
problem have made use of this important work.  However, in the present
paper we show that the solutions may take  more general forms than
those found by Goodman. For the sake of completeness we reproduce
Goodman's arguments in Appendix \ref{sec:AppAA} and indicate  the point
at which we depart from them.

In section \ref{bog} we construct and solve the mean field
approximation, starting from a number- and isospin-conserving
hamiltonian, and applying an appropriate transformation from particles
to quasi-particles. As a result of such transformation the proton
$\Delta_p$, the neutron $\Delta_n$ and the isoscalar $\Delta_0$ gaps
may take non-vanishing values.  We look for non-trivial solutions in
which the three of them are different from zero. Although the formalism
is at least valid for any situation involving separable pairing
interactions, we restrict the application to nuclei with equal number
of protons and neutrons. The nature of the solutions thus obtained is
discussed in detail for a single $l$-shell (section \ref{singlel}) and
for a single $j$-shell (section \ref{singlej}). A comparison between
exact and approximate results is made in the former case.  Since
significant differences may be found in the behavior of the solutions
for spin saturated and unsaturated systems, a suggestion is made in the
Conclusions (section \ref{con}) about the regions of the nuclear chart
in which non-trivial solutions may be favored.

In the main part of the paper we include only general arguments.
All the details of the calculations can be found in the Appendices
\ref{sec:AppBB}, \ref{sec:AppCC} and \ref{sec:AppDD}.

In the present paper we confine the discussion to the problem of the
deformed mean field treatment. Improvements over this approximation
involves the relation between the laboratory system, and the instrinsic
system in which the deformation takes place. Projection techniques have
recently been used to this purpose\cite{DP98}.

\section{The generalized BCS treatment}   \label{bog}
\subsection{The transformation}    \label{big}

Let us consider a general, number- and isospin-conserving hamiltonian.
Assuming that it may be treated within a (gauge- and isospin-deformed)
field approximation, we define the following transformation to
quasi-particles:
\beqa
\left(\begin{array}{r} \alpha^{\dag}_{p\lambda \mu}\\ 
\alpha^{\dag}_{n\lambda \mu}
\end{array}\right)&=&
\left(\begin{array}{cc} u_{pp}&iu_{np}\\
-iu_{pn}&u_{nn}\end{array}\right)
\left(\begin{array}{r}
 c^{\dag}_{p\lambda \mu}\\c^{\dag}_{n\lambda \mu}\end{array}
\right) 
\,+\,\left(\begin{array}{cc} -v_{pp}&-iv_{np}\\
iv_{pn}& -v_{nn}\end{array}\right)
\left(\begin{array}{r} c_{p\lambda{\bar \mu}}\\c_{n\lambda{\bar \mu}}\end{array}\right) ,
\label{genbog}
\eeqa
with $u_{vw},v_{vw}$ real, where $v=$(proton, neutron)$=(p,n)=(1,- 1)$;
$\lambda$ denotes  either $j$ or
$(l,s)$,  
while $\mu$ labels the corresponding projections\footnote{More details
of the notation are given in Appendix \ref{sec:AppBB}.}($\mu>0$ in
(\ref{genbog})). The operator $c^{\dag}_{v\lambda {\bar \mu}}$ creates a
particle in the time-reversed state of $c^{\dag}_{v\lambda \mu}$.  The
time-reversed and hermitean-conjugate versions of (\ref{genbog})
complete an 8 $\times$ 8 transformation, which reduces to two 4
$\times$ 4 ones.

The transformation (\ref{genbog}) simplifies over the one given
in ref. \cite{go72} on two accounts:\\
i) It does not mix operators creating (or annihilating) 
particles in time-reversed states. As shown by Goodman,
this pairing only becomes relevant for well shape-deformed
nuclei.\\
ii) The transformation (\ref{genbog}) only yields non-vanishing
expectation values  for the components $J=0,T=1,T_0=v=\pm 1$ and $J=1,
J_0=T=0$ of the tensor $(S^{JT}_{J_0T_0})^{\dag}$ constructed as a product
of two single-particle creation operators coupled to spin $J$ and
isospin $T$, with projections $J_0$ and $T_0$, respectively.  In spite
of this limitation, the transformation  takes into account the full
neutron-proton pairing. The case is completely analogous to the
familiar situation in shape-deformed nuclei, in which the expectation
values of the quadrupole tensor are:
\beqa
<Q_0> \neq 0 \;\;\;\; &,& \;\;\; <(Q_2+Q_{-2})> \neq 0 ,\non \\
<Q_{\pm 1}>&=&<(Q_2-Q_{-2})> = 0 . \label{intr}
\eeqa
The first line, non-vanishing expectation values  are the (large)
``order parameters'' of the problem, while the ones in the second line
define the intrinsic system.  Equations (\ref{intr}) violate the
rotational invariance in the  intrinsic system but not in the
laboratory system. We may restore such invariance through the
introduction of collective coordinates, within an appropriate formalism
(see, for instance, Ref.\cite{bk90}).  Some of the possible order
parameters may vanish if there is a subsymmetry still conserved, like
for an axially symmetric deformation.  As a consequence, the Nilsson
model takes into account the full, spherically symmetric, quadrupole
interaction within the mean field approximation.

In a completely similar way, the transformation (\ref{genbog}) may be
used to treat neutron-proton pairing problems.  In particular, we do
not need to include any real non-diagonal components in the
transformation (\ref{genbog}), which would yield a non vanishing
expectation value for the component $T=1,T_0=J=0$, since this component
plays a similar  role as those in the second line of (\ref{intr}).  The
fact that in the deformed field treatment of an isovector pairing force
there are only two order parameters and that, through a convenient
orientation of the intrinsic system in gauge- and isospace, such
parameters may be chosen as the $nn$ and $pp$ gaps (orientation A) is
demonstrated in Refs.\cite{GW68} and \cite{DU71}. Other (equivalent)
orientations of the intrinsic system are possible such as the one in
which the two order parameters are represented by the value of the $np$
gap and the equally valued $pp$ and $nn$ gaps (orientation B, cf.
Ref.\cite{DU71}). Again, the introduction of an additional subsymmetry
between protons and neutrons may eliminate one of the two parameters:
either the difference between the $nn$ and $pp$ gaps in case of
orientation A or the $nn$ and $pp$ gaps in orientation B.  This is
consistent with the results obtained in Refs.\cite{En96} and
\cite{En97} for nuclei with equal number of protons and neutrons, that
there are two BCS-like solutions with $T=1$ pairs: one with no $np$
pairs and the other composed entirely by $np$ pairs. Both are
degenerate and do not mix with each other. We argue that they
correspond to the same state described from different intrinsic
sytems.  Obviously, the physical results obtained are independent of
the (unphysical) orientations of the intrinsic system, if a proper
treatment is applied for relating the values of the magnitudes in the
intrinsic and laboratory frames of reference\footnote{The formalism of
Ref.\cite{bk90} has been recently applied to the isospin case
\cite{bcs99}.}.

The orthonormalization conditions are:
\beqa
\sum_s(u_{sv}u_{sw}+v_{sv}v_{sw})=\delta_{vw}\;;\;\;\;\;\;
\sum_s(u_{sv}v_{sw}-u_{sw}v_{sv})=0 .
\eeqa

In the present paper we use the transformation (\ref{genbog}) 
in order to construct the independent quasi-particle hamiltonian.
We also study  properties of the vacuum state $|>$, which is annihilated
by the quasi-particle annihilation operators $\alpha_{v\lambda\mu}$,
i.e.,
\beqa
|>&=& \Pi_{\lambda, \mu>0}\bigl[ u_{pp}u_{nn}-u_{pn}u_{np}\non \\
&&+(u_{nn}v_{pp}-u_{np}v_{pn})
c^{\dag}_{p\lambda\mu}c^{\dag}_{p\lambda{\bar \mu}} 
+(u_{pp}v_{nn}-u_{pn}v_{np})
c^{\dag}_{n\lambda\mu}c^{\dag}_{n\lambda{\bar \mu}}\non \\
&&+i(u_{nn}v_{np}-u_{np}v_{nn})
(c^{\dag}_{n\lambda\mu}c^{\dag}_{p\lambda{\bar \mu}}
-c^{\dag}_{p\lambda\mu}c^{\dag}_{n\lambda{\bar \mu}})\non \\
&&+(v_{pp}v_{nn}-v_{np}v_{pn})
c^{\dag}_{n\lambda\mu}c^{\dag}_{n\lambda{\bar \mu}}
c^{\dag}_{p\lambda\mu}c^{\dag}_{p\lambda{\bar \mu}}\bigr]
|\mbox{vacuum}> .
\eeqa

In spite of the fact that the calculation is made in the intrinsic
system, neither the vacuum energy nor the quasi-particle excitation
energies are modified by the corrections needed to restore the
symmetries (at least to leading order, within an expansion on the
inverse of the order parameters).

\subsection{The  solution} \label{sol}

In order to keep the formalism as simple as possible,
we consider in this work a separable pairing Hamiltonian of the form
\beqa
H&=& \sum_{v \lambda}e_{v\lambda}
\sum_{\mu}c^{\dag}_{v\lambda \mu}c_{v\lambda \mu}           
\non \\
&&\,-\,g_1\sum_{T_0}(S^{01}_{0T_0})^{\dag}S^{01}_{0T_0}
\,-\,g_0\sum_{J_0}(S^{10}_{J_00})^{\dag}S^{10}_{J_00} .
\label{ham}
\eeqa

It is useful to parametrize the interaction strengths as
\beqa
g_1=g\,(1-x)\;;\;\;\;\;\;g_0=g\,(1+x)\;;\;\;\;\;\;
-1\leq x \leq  1 . \label{x}
\eeqa 

The matrix elements of the transformation (\ref{genbog}) are fixed
through the following steps:\\
i) The diagonalization of the 
two-quasi-particle Hamiltonian
\beqa
H^{(11)}&=&
\sum_{v,\lambda}\epsilon_{v\lambda} 
\sum_{\mu}(c^{\dag}_{v\lambda \mu}c_{v\lambda \mu})^{(11)}
    \non \\
&&-\sum_v\Delta_v(S^{\dag}_v+S_v)^{(11)}
\,-\,\Delta_0 (S^{\dag}_0+S_0)^{(11)} , \label{qp}
\eeqa
where the supraindex $(11)$ denotes as usual the product 
of a quasi-particle creation  and an annihilation operator.
Using the shorthand notation $S_v=S^{01}_{0v}$ and $S_0=
S^{10}_{00}$, the gaps are defined as $\Delta_v=g_1<S_v>$ and $\Delta_0=g_0<S_0>$. The single-particle energies $\epsilon_{v\lambda}=
e_{v\lambda}-\lambda_v$ include a Lagrange multiplier 
$\lambda_v$. The diagonalization of the hamiltonian
(\ref{qp}) is equivalent to the matrix diagonalization 
(cf. Appendix \ref{sec:AppBB})
\beqa
\left(\begin{tabular}{cccc}
$\epsilon_{p\lambda}$ & 0 &$\Delta_p$&$k_{\lambda \mu}\Delta_0 $\\
0&$\epsilon_{n\lambda}$ & $k_{\lambda \mu}\Delta_0$& $\Delta_n$ \\
$\Delta_p$&$k_{\lambda \mu}\Delta_0$ &$-\epsilon_{p\lambda}$&0\\
$k_{\lambda \mu}\Delta_0$&$\Delta_n$ & 0&$-\epsilon_{n\lambda}$
\end{tabular}\right) 
\left(\begin{tabular}{c} $u_{pv}$\\$u_{nv}$\\$v_{pv}$\\$v_{nv}$
\end{tabular}\right)=E_{v\lambda\mu}\left(\begin{tabular}{c} 
$u_{pv}$\\$u_{nv}$\\$v_{pv}$\\$v_{nv}$
\end{tabular}\right)  \label{mat}
\eeqa
that
yields the quasi-particle energies $E_{v\lambda\mu}$ and the
coefficients $u_{vw},v_{vw}$ as functions of the single-particle
energies and gap parameters. The factors
$k_{\lambda\mu}$ are defined in Table \ref{t1}.\\
ii) The fulfillment of the self-consistency conditions [cf.
Eq.~(\ref{eq:B2})]:
\beqa
\Delta_v=g_1 \sum_{w,\lambda,\mu>0} u_{vw} v_{vw}\;;
\;\;\;\;\;\;
\Delta_0=g_0 \sum_{v,\lambda,\mu>0} k_{\lambda \mu} 
(u_{pv} v_{nv} + u_{nv} v_{pv}). \label{self}
\eeqa
iii) The fulfillment of the number equations [cf. Eq.~(\ref{eq:B2})]:
\beqa
N_v=\sum_{w,\lambda,\mu} v^2_{vw}. \label{lagr}
\eeqa

The vacuum energy is given by the expectation value ($<H>$):
\beqa
W_{gs}&=&\sum_{v,\lambda}e_{v\lambda}\sum_{w,\mu} v^2_{vw}\;-\;
\frac{\Delta^2_p+\Delta^2_n}{g(1-x)}\;-\;\frac{\Delta_0^2}{g(1+x)} .
\label{wgs}
\eeqa

In this paper we are interested in situations with the same number
of protons and neutrons. If we further assume the same
single-particle spectrum for both kind of particles
$(e_{p\lambda}=e_{n\lambda})$, the symmetry of the problem 
requires $\Delta_p^2=\Delta_n^2$ but leaves open the relative
sign of $\Delta_p$ and $\Delta_n$. Therefore
we expect two non-trivial solutions with
$\Delta_p=\pm \Delta_n\neq 0$, $\Delta_0\neq 0$  and two
trivial solutions\footnote{The relative sign of $\Delta_p$ and 
$\Delta_n$ is immaterial for the two trivial solutions.} 
with $\Delta_p=\Delta_n=0$, $\Delta_0\neq 0$
and $\Delta_p=\Delta_n\neq0$, $\Delta_0= 0$.

To explore the characteristics of these solutions, in what follows we
work in a model space consisting of one single shell. In this case, the
single-particle energies $e_{v\lambda}$ disappear from the problem and
all energies become simply proportional to the strength
parameter\footnote{We take $g=\frac{1}{2}$ in the numerical
calculations below.}.

We have the option to work with an $l$-shell or a $j$-shell.
There are at least two advantages for considering the case of a $l$-
over a $j$-shell: i) The pair of particles may couple to orbital
angular momentum $L=0$, even for $J=1$. ii) The pairing problem  is
amenable to a numerical solution and we may check the BCS results
against the exact ones. Nevertheless, we also carry the calculation
for the case of a single $j$-shell, in order to study the effect
of the pairing hamiltonian in a spin non-saturated system. 

\section{The case of a single $\lowercase{l}$-shell} \label{singlel}

\subsection{The BCS solution}

The solution with $\Delta_n=- \Delta_p$  is the one allowed by Goodman
but for the fact that the transformation to quasi-particles displays
non-diagonal coefficients $u_{vw} \neq 0$ [Eq.~(\ref{uvg})].  However,
since the eigenvalues of the Hamiltonian are degenerate, it is always
possible to make linear combinations of the solutions corresponding to
the quasi-protons and quasi-neutrons (i.e., to the two values of $v$)
and impose the condition $u_{np}=u_{pn}=0$ [cf. Eq.~(\ref{mat})].

However there is a sharp limitation in the domain of validity of this
solution given by the self-consistency condition [Eq.~(\ref{self-})],
which implies $x=0$ [(cf.  Eq.~(\ref{x}))]. Moreover, the allowed
interval around this value has zero width.

\begin{figure}[p]
\centerline{\psfig{figure=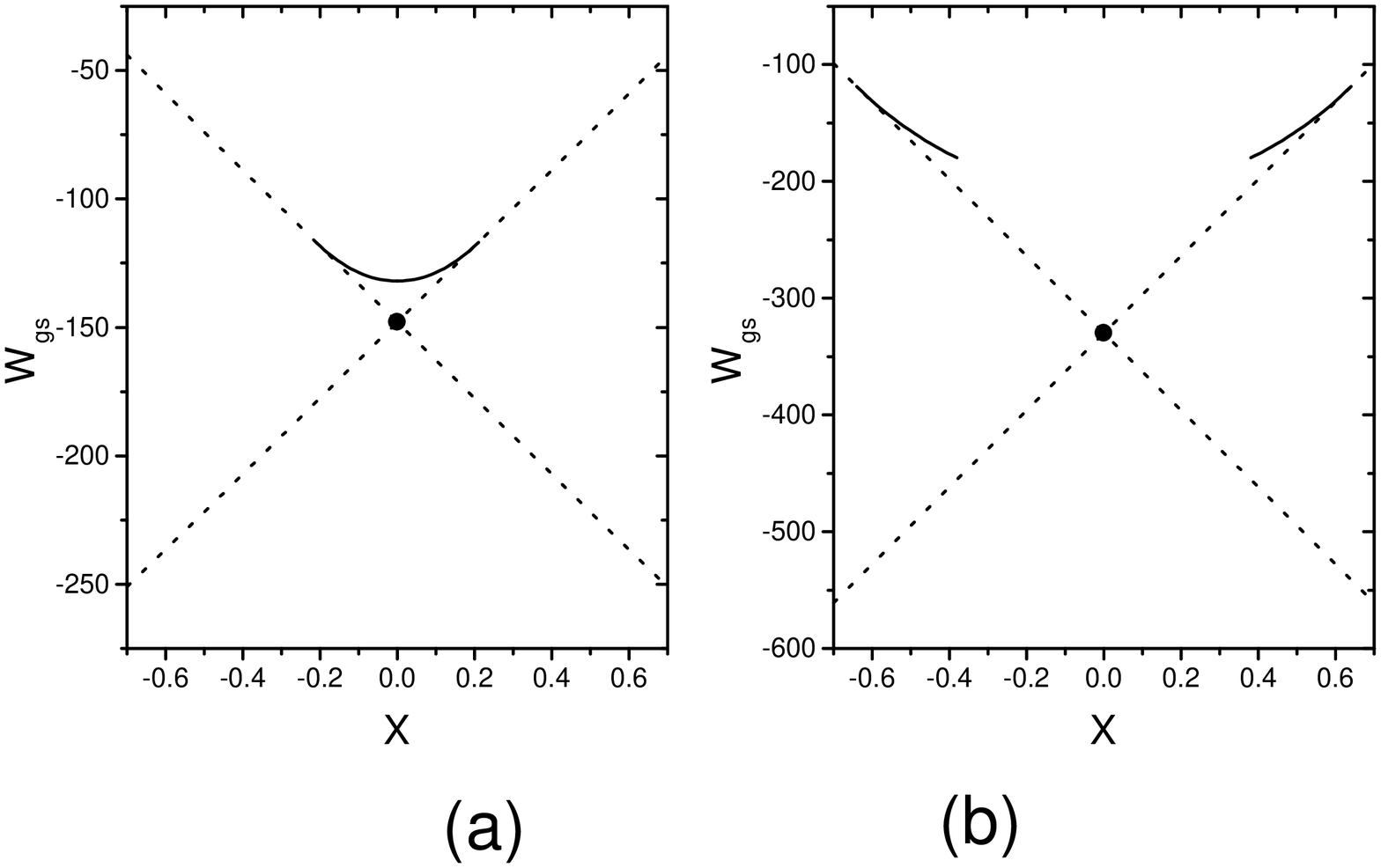,height=11.0cm}}
\caption[]{The vacuum energies $W_{gs}$ for a single $l$-shell as a
function of $x$, (a) for $\eta=-0.80$ and (b) $\eta=-0.46$.  Dotted
lines: trivial  solutions. Full:  solution with $\Delta_p=\Delta_n$.
Big dot: solution with $\Delta_n=-\Delta_p$.}
\label{f1} 
\end{figure}

Figure \ref{f1} represents the vacuum energies as a function of $x$ for
two different values of the occupation parameter $\eta$ (see Table
\ref{t1}).  The energy $W_{gs}$ associated with the solution having
$\Delta_p=-\Delta_n$ lies exactly at the crossing point of the energies
corresponding to the two trivial solutions ($\Delta_0=0$ or
$\Delta_p=0$). Since the solution given in (\ref{cara}) only determines
the total gap $\sqrt{\Delta_p^2+\frac{1}{2} \Delta_0^2}$, but not the
ratio between the isovector and isoscalar gaps, the existence of this
solution  may be interpreted (in the present case) as a manifestation
of the degeneracy associated with the crossing,  without further
physical meaning.

The two trivial solutions exist for any value of the parameter $x$.

The solution with $\Delta_n=\Delta_p$ requires the numerical treatment
of two simultaneous equations [i.e.,~(\ref{num}) and (\ref{selfx})],
depending on the two variables
\beqa
\zeta=\epsilon/\Delta_p\;\;\; \mbox{and} 
\;\;\;\gamma=\Delta_0/\Delta_p .
\label{horn} 
\eeqa 
The task is simplified by the existence of transformations that leave
invariant the system of equations [Eqs.~(\ref{inv1}) and
(\ref{inv2})].  Thus it is sufficient to solve the system 
in one quadrant of the cartesian system determined by  the strength
parameter $x$ and the occupation parameter $\eta$.

In Fig.~\ref{f1}(a) there appears an upper value of $|x|$ for which
there is a solution. In Fig.~\ref{f1}(b) there is also a lower value of
$|x|$.  In the following we explain this behavior.

According to Fig.~\ref{f1}, the vacuum energy $W_{gs}$ of the solution
with $\Delta_p=\Delta_n$ lies always higher than the energies of the
trivial solutions and approaches these preceding values at the extremes
of the domain in which it exists. For instance, for negative values of
$x$, the energy of the solution with $\Delta_p=\Delta_n$ becomes the
energy of the highest trivial solution, namely the one  with a
non-vanishing isoscalar gap.  This fact suggests that one limit of the
domain in the $(x,\eta)$ plane is obtained from solutions such that
$\gamma\rightarrow \infty$ and $\zeta\rightarrow \infty$.  In such a
limiting case the following curve can be found [cf. Eqs.~(\ref{infy})]:
\beqa
\eta_1&=& \sqrt{\frac{1+x_1}{1-x_1}}\;;\;\;\;\; 
(-1\leq x_1\leq 0) . \label{panda} 
\eeqa
The procedure  also yields the (vanishing) value of $\Delta_p$; the
(non-vanishing) value of $\Delta_0$; and the vacuum energy $W_{gs}$ on
this curve. We may verify that they are equal to the results of the
trivial solution with $\Delta_p=0$ that are given in Table \ref{t2}.

The limiting curve for positive values of $x_1$ is easily obtained from
(\ref{panda}) by means of the transformation (\ref{inv1}) (Note that
in this case the gap $\Delta_p \rightarrow {\Delta_0}/{\sqrt{2}}$
and $\Delta_0\rightarrow 0$ [cf. Eq.~(\ref{sire})].)
\beqa
\eta_1&=& \sqrt{\frac{1-x_1}{1+x_1}}\;;\;\;\;\; 
(0\leq x_1\leq 1) . \label{pandap} 
\eeqa

\begin{figure}[p]
\centerline{\psfig{figure=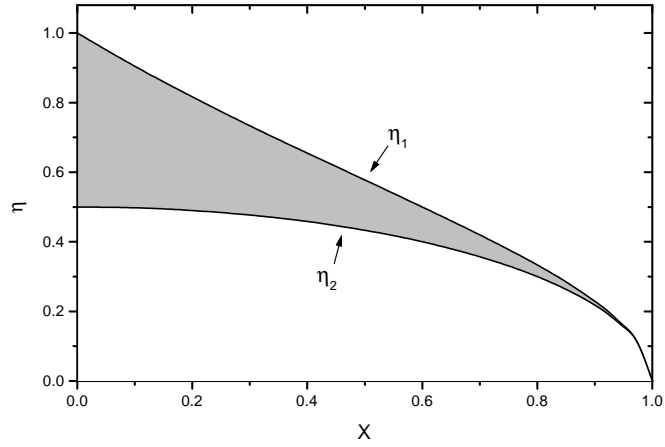,height=7.0cm}}
\caption[]{The allowed region  in the $(x,\eta)$ plane for
$\Delta_n=\Delta_p$.}
\label{f3} 
\end{figure}

The curve (\ref{pandap}) that limits the allowed region is shown 
in Fig.~\ref{f3}.  

According to Fig.~\ref{f1}, the solution with $\Delta_p=\Delta_n$ does
exist for $\eta=-0.80$ in an interval around $x=0$,  but it does not
exist for $\eta=-.46$ in an  interval enclosing the same point.  In
order to explain this behavior we construct a second  boundary for the
domain of validity by studying this solution in the limit
$\zeta\rightarrow 0$ and $\gamma \rightarrow \sqrt{2}$. The locus at
which this condition is satisfied in the ($x,\eta$) plane is [cf.
Eqs.~(\ref{cebra})]:
\beqa
\eta_2= \frac{1}{2}\sqrt{1-x^2_2} . \label{oso}
\eeqa
Equation~(\ref{oso}) indicates that the curve does not exist for values of
the occupation parameter $\eta \geq \frac{1}{2}$. Therefore, for such
values of $\eta$, the region of validity is only limited by Eqs.
(\ref{panda}) and (\ref{pandap}).  For values $\eta \leq \frac{1}{2}$,
the allowed solutions lie in the interval $x_2\leq x\leq x_1$. This
interval gets narrower as we approach the half-filled shell and, in
fact, it vanishes in the limits $x_1 \rightarrow x_2 \rightarrow 1$
(Fig.~\ref{f3}). These predictions are consistent with the numerical
results presented in Fig.~\ref{f1}.

The absence of solutions within the interval  $0\leq x\leq x_2$ may be
interpreted as due to the presence of  unstabilities in the generalized
BCS solution:  the quasi-neutron energy $E_n$ vanishes along the curve
(\ref{oso}).

\subsection{The comparison with the exact results}

The Hamiltonian of Eq.(\ref{ham}) is invariant under the
transformations of the group $SO(8)$. Thus it is solvable by an
appropriate choice of the infinitesimal operators\cite{pa69,edmp81}.
The eigenstates $|k>$ of the Hamiltonian can be expressed in terms  of
a basis labeled by a chain of irreducible representations of $SO(8)$
and its subgroups, namely,
\begin{equation}
|k;S M_S, T M_T> = \sum_n c^{(k)}_{n}(SM_STM_T)
	| \Omega,\lambda,[n,n,0],SM_S,TM_T > ,
\end{equation}
where $\Omega=2l+1$ labels the representations of $SO(8)$ for the case
of shell seniority zero, $\lambda = \Omega- ({A}/{2})$ ($A=N_p+N_n$)
is related to the representations of $SO(7)$, while $[n,n,0]$ labels
the representations of the group $SU(4)$ which is homomorphic to
$SO(6)$.  The group $SU(4)$ can be readily decomposed in $SU(2) \times
SU(2)$ so that the quantum numbers $S$, $M_S$, $T$ and $M_T$ are
sufficient to completely specify the states of $SO(6)$.

\begin{figure}[p]
\centerline{\psfig{figure=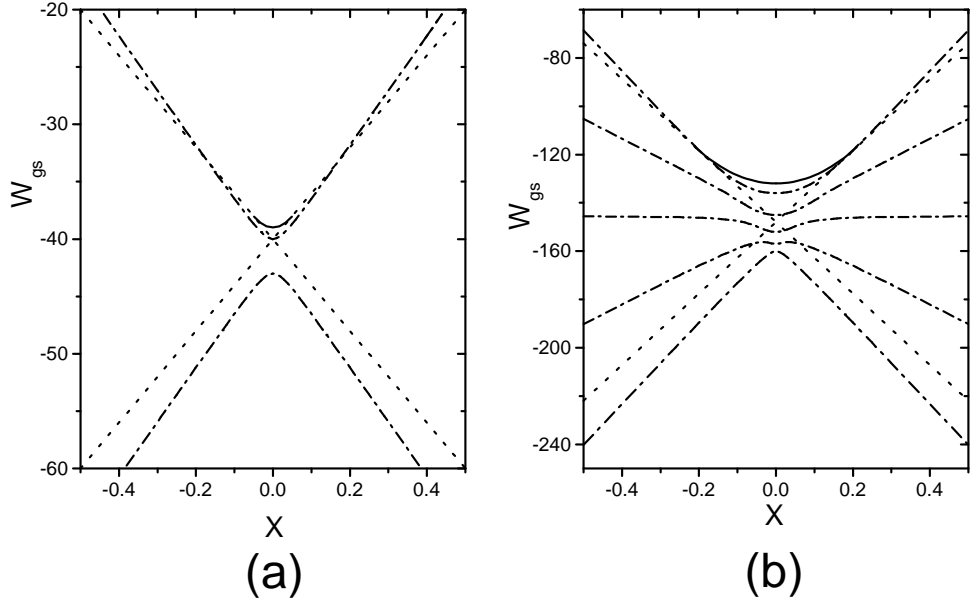,height=11.0cm}}
\caption[]{The energy spectrum as a function of $x$ for $\Omega=41$.
(a) $A=4$ and (b) $A=16$.  Dot-dashed lines: exact solutions. Dotted:
trivial solutions. Full:  solution with $\Delta_n=\Delta_p$.}
\label{exact1} 
\end{figure}

We have obtained the solutions of the Hamiltonian as a function of the
parameter $x$ introduced in (\ref{x}). The corresponding spectra for
$S=0$ and $T=0$ are shown in Fig.~\ref{exact1}. The value $l=20$ has 
been used. It is sufficiently large to allow for a clear 
manifestation of collective effects.

If $A=4<<\Omega$ ($\eta\approx -1$), the exact solution displays
only two eigenvalues. The
two trivial solutions, $\Delta_0=0$ or $\Delta_p=0$,
correspond to the lowest eigenvalue for $x$ smaller and larger than
zero, respectively [Fig.\ref{exact1}(a)]. The approximate solution is
somewhat less bound than the exact one, as to be expected. The correspondence is inverted
for the highest eigenvalue. However there is a small interval around
$x\approx 0$ in which the exact eigenvalues display some curvature.
While the highest eigenvalue may be correlated with the solution with
$\Delta_p=\Delta_n$, no such correlation can be made for
the solution  $\Delta_p=-\Delta_n$  in the case of the
lowest state.

For a larger number of particles ($A=16, \eta=-0.80$), there appear
more eigenvalues [Fig.~\ref{exact1}(b)]. However, as in
Fig.~\ref{exact1}(a), there is a correspondence between the trivial
solutions and the lowest and highest eigenvalues, at least for values
of $|x|>0.2$.  Moreover, for this case, a  strong correlation can be made
between the eigenstate with the highest eigenvalue and the solution with
$\Delta_p=\Delta_n$ throughout the interval $-0.2 <x<0.2$. On the
contrary, the lowest eigenvalue continues to be represented by two
straight lines with a relatively sharp crossing at $x \approx 0$, which
is consistent with the absence of a solution with
$\Delta_p=-\Delta_n$.

\section{The case of a single $\lowercase{j}$-shell} \label{singlej}

In order to solve the self-consistency and number equations
[(\ref{hiena}), (\ref{sc}) and (\ref{chac})] we must select a value for
the quantum number $j$. The value $j=\frac{21}{2}$ is large enough to
allow for a clear display of collective effects.

\begin{figure}[p]
\centerline{\psfig{figure=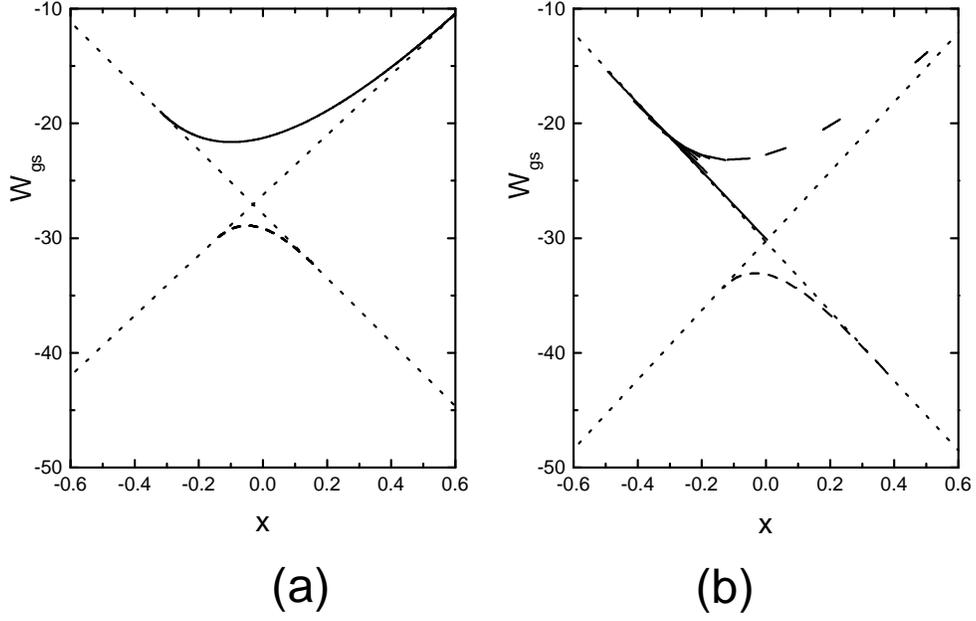,height=11.0cm}}
\caption[]{The vacuum energies $W_{gs}$ for a single $j$-shell as a
function of $x$, for (a) $\eta=-4/11$ and (b) $\eta=0$.  Dotted lines:
trivial solutions. Dashed: solution with $\Delta_n=-\Delta_p$. Full:
solution  $\Delta_n=\Delta_p$.}
\label{f4} 
\end{figure}

The vacuum energy $W_{gs}$ of the two non-trivial and of the two
trivial solutions are represented as a function of the strength
parameter $x$ in Fig.~\ref{f4}. The occupation parameters $\eta$ (see
Table \ref{t1}) are taken to be $\eta$=-4/11 and 0 in Figs.~\ref{f4}
(a) and (b), respectively.

At variance with the case of a single $l$-shell, the solution with
$\Delta_p=-\Delta_n$ exists within a finite domain for comparable
isovector and isoscalar strengths ($x\approx 0$). Its vacuum energy
becomes the lowest energy within this domain. At the extremes of the
allowed interval, Goodman's solution merges with the most favored
trivial solution corresponding to the same value of the parameter $x$.
This behavior is common to the two values of the occupation parameter.

The non-trivial solution with $\Delta_p=\Delta_n$ also exists for the
two occupation parameters. As similarly to the case of a $l$-shell,
this solution always has the highest energy. The $x$ domain of
existence is somewhat broader than the one associated with the solution
for $\Delta_p=-\Delta_n$. At each extreme of the allowed interval the
solution merges with the most unfavored trivial one corresponding to
the same value of $x$.

According to Fig.~\ref{f4}, the vacuum energy of the solution with
$\Delta_p=\Delta_n$ is a continuous function of $x$ for $\eta=4/11$ and
a discontinuous one for $\eta=0$ (half-filled shell). In the second
case the energy only exists for certain intervals of $x$ and is a
linear function of $x$ within each interval.

\begin{figure}[p]
\centerline{\psfig{figure=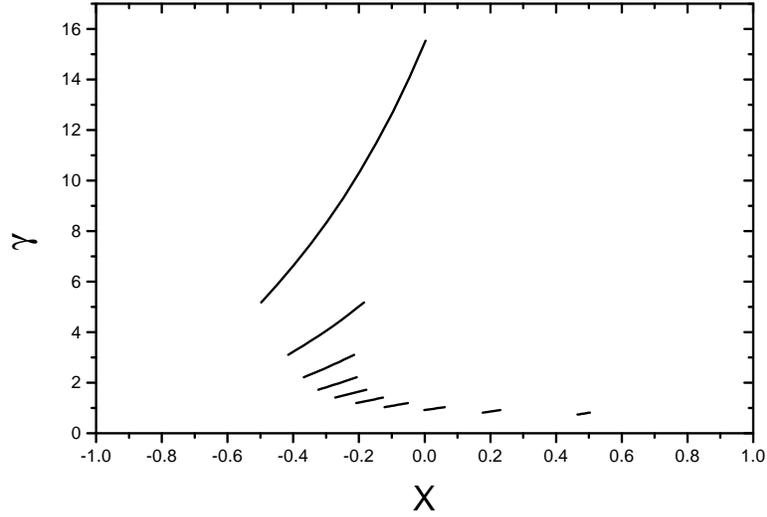,height=8.0cm}}
\caption[]{The asymmetry parameter $\gamma$ as a function of $x$, for
$\eta=0$ (solution with $\Delta_n=\Delta_p$).}
\label{f5} 
\end{figure}

The  case $\eta=0$ is treated analytically in Appendix \ref{roto2}.
Since the Lagrange multipliers vanishes, the number equation is
trivially fulfilled.  The discontinuities may be traced back to the
vanishing of the quasi-neutron energies $E_n$ for certain values of the
asymmetry parameter $\gamma$ (Eq.~(\ref{jil})). Such discontinuities
determine closed intervals within which the self-consistent conditions
yield the parameter $x$ as a continuous function of $\gamma$.  Within
each interval, $\gamma$ lies between the roots corresponding to two
consecutive values of $m$ [Eq.~(\ref{disc})] . The intervals may be
labeled by the value of $k=m+\um$, where $m$ is the smallest magnetic
quantum number. There are $j-\frac{1}{2}$ closed intervals. The
function $x(\gamma)$ may be inverted and thus an analytical function
$\gamma(x)$ is obtained within each interval [Fig.~\ref{f5} and
Eq.~(\ref{card})]. Note  that there are several values of $\gamma$
which are compatible with a single value of $x<0$ . The opposite
situation appears for $x>0$, where there may not be any value of
$\gamma$ associated with certain values of $x$.

The vacuum energies are given in  (Eq.~\ref{gorr}).  Within each
interval $k$, they confirm the linearity with $x$  that was found in
the numerical solution displayed in Fig.~\ref{f4}. The slope of the
lines $W_{gs}(x)$ increases with the value of $k=m+\frac{1}{2}$. In the
first interval the solution is degenerate with the trivial solution
$\Delta_0 \neq 0$ and extends within the interval $-\frac{1}{2}\leq x
\leq 0$ (to leading order in $1/j$).

We also obtain an expression for the extremes $x_{\pm}$ limiting each
interval (Eq.~\ref{pal}).

In addition to the  $j-\frac{1}{2}$ closed intervals, there are also
open intervals for small and large values of $\gamma$.  For $\gamma <
{1}/{\sqrt{2j(j+1)}}$ and $\gamma > \sqrt{{j}/{[2(j+1)]}}$ the
solutions collapse to the points $x=\mp 1$, respectively, and the
energy vanishes.

\section{Conclusions} \label{con}

We have studied a generalization of the BCS
approximation that takes into account both the isoscalar and isovector
pairing. Although for separable forces the solution is always amenable
to the diagonalization of a $4 \times 4$ matrix equation, in this paper
we have only considered the case of identical number of protons and
neutrons and  identical single-particle energy levels for both kind of
nucleons. We perform the calculation in an intrinsic system in which
the non-vanishing gaps are the proton gap $\Delta_p$, the neutron gap
$\Delta_n$ and the isoscalar gap $\Delta_0$. We label the solution as
non-trivial if none of the three gaps vanishes and as trivial if either
the isoscalar gap or the neutron and proton gaps  equal zero. We expect
two non-trivial solutions with $\Delta_p=\pm \Delta_n$ and two trivial
solutions. The non-trivial solution with the minus sign corresponds to
the one allowed by Goodman's argumentation.

We have studied the vacuum energy of the four solutions in the case of
a  single $l$- and a single $j$-shell.  We found that the two
non-trivial solutions behave differently in spin-saturated than in
spin-non-saturated systems.

In the case of a $l$-shell, the solution with $\Delta_p=-\Delta_n$ does
not exist. More precisely, it only exist for the value $x=0$ of the
strength parameters, within an interval of zero width (Fig. \ref{f1}).
The vacuum energies of the two trivial solutions also cross at this
point and have the same value as the non-trivial one. The exact
calculation confirms the non-existence of a solution with
$\Delta_p=-\Delta_n$, since the lowest energy is well represented by
two straight lines crossing rather sharply at $x=0$.

On the contrary, the solution with $\Delta_p=-\Delta_n$ exists within a
finite interval in the case of a $j$-shell. In this interval, starting
at $x\geq -1/7$, it becomes the lowest state of the system. As $x$
tends to $-1/7$ from above, the non-trivial solution tends to the
trivial one with $\Delta_0=0$, which becomes the lowest state in the
interval $-1\leq x \leq -1/7$. The non-trivial solution tends to the
trivial solution with an isoscalar gap at the other extreme of its
region of existence.

The solution with $\Delta_p=\Delta_n$ exists and has similar properties
for both kind of shells: it always displays the highest vacuum energy
within its regions of existence.  For sufficiently large values of
$|x|$ it merges with the highest trivial solution. In the case of an
$l$-shell, we obtain expressions for its domain of existence in the
plane determined by the strength $x$ and the occupation number $\eta$.
The energy of the quasi-neutrons vanishes and the system becomes
unstable along one of the two curves limiting this domain. The pattern
is confirmed by checking the BCS results against exact calculations: it
is possible to identify the highest exact eigenvalue with this
non-trivial solution.  This general picture is very similar for the
case of a $j$-shell.  The main difference appears for a half-occupied
shell ($\eta=0$), where the regions of validity consist of successive
intervals which properties may be predicted analytically.

Concerning the observation of the effects discussed in the text we may
conclude that the experimental search for (large) non-trivial solutions
should focus on nuclei such that both valence nucleons mainly occupy
the $j$-level which becomes integrated to the lower harmonic oscillator
shell.

\acknowledgments

This work was supported in part by Fundaci\'on Antorchas, the Consejo
Nacional de Investigaciones Cient{\'\i}ficas y T\'ecnicas under
Contracts PIP 4004/96 and PIP 4486/96 and the Agencia Nacional de
Promoci\'on Cient{\'\i}fica y Tecnol\'ogica of Argentina under contract
PICT03-00000-00133.

\newpage
\appendix
\section{Some symmetry considerations} \label{sec:AppAA}

In order to reach the point where our results depart from Goodman's, in
this Appendix we follow its presentation of the consequences of the
symmetries involved\cite{go72}.  In Goodman's notation, the
transformation (\ref{genbog}) reads:
\beqa
\left(\begin{array}{l} \alpha^{\dag}\\ \alpha\end{array} \right) 
&=&\left(\begin{array}{cc} U &-V\\ -V^* & U^*\end{array}
\right) \left(\begin{array}{l} c^{\dag}\\c\end{array} \right) ,
\label{charlotte}
\eeqa
where $\left(c^{\dag}\right)$ = $
\left(c^{\dag}_{p\lambda \mu},c^{\dag}_{n\lambda\mu}, 
c^{\dag}_{p\lambda {\bar \mu}},c^{\dag}_{n\lambda{\bar \mu}}\right)$
and
\beqa 
U = \left(\begin{array}{cccc} u_{pp} & iu_{np}&0&0\\
-iu_{pn}&u_{nn}& 0 & 0\\
0&0&u_{pp}&-iu_{np}\\ 0&0&iu_{pn}&u_{nn}\end{array}\right) ,
\;\;\;\; V = \left(\begin{array}{cccc}
0&0&v_{pp}&iv_{np}\\ 0&0&-iv_{pn}& v_{nn}\\
-v_{pp}&iv_{np}&0&0\\ -iv_{pn}&-v_{nn}&0&0\end{array}\right) . 
\label{cosak}
\eeqa
 
The generalized density  matrix is defined as
\beqa
{\cal R}&=&\left(\begin{array}{cc} 
\rho & t\\t^{\dag}& 1-{\tilde \rho} \end{array}\right) ,
\non \\
\rho_{vw}&= & <c^{\dag}_{wjm}c_{vjm}> = 
(V^*\,{\tilde V})_{vw} , \non \\
t_{vw}&= & <c_{wjm}c_{vjm}>=(UV^{\dag})_{vw} ,
\eeqa
where transpose matrices are indicated by a tilde and the $\rho$ and
$t$ matrices satisfy the relations
\beqa
\rho=\rho^{\dag}\;;\;\;\;\;\;\;{\tilde t}=-t . \label{rest1}
\eeqa
The requirement ${\cal R}={\cal R}^2$ is equivalent to: 
\beqa
\rho-\rho^2 = t\,t^{\dag}\;;\;\;\;\;\;\;\rho \,t = t\,{\tilde \rho} . 
\label{rest2}
\eeqa

Time reversal conservation implies the existence of  relations between
the matrix elements of $\rho$ and $t$. If in addition the restrictions
(\ref{rest1}) are applied, one obtains $\rho$ and $t$ matrices of the
form
\beqa
\rho = \left( \begin{array}{cccc} 
\rho_{pp}& \rho_{pn}& 0 &0\\
\rho_{pn}^* & \rho_{nn}&0 &0 \\
0&0 & \rho_{pp} & \rho_{pn}^* \\
0&0 & \rho_{pn} & \rho_{nn} \end{array} \right) , 
\;\;\;\; t = \left(\begin{array}{cccc}
0&0& t_{pp}& t_{pn}\\
0&0& t_{pn}^*& t_{nn}\\
-t_{pp}&-t_{pn}^*&0&0\\
-t_{pn} & -t_{nn}& 0 & 0 \end{array} \right) , \label{makine}  
\eeqa
with the diagonal terms $\rho_{pp}, \rho_{nn}, t_{pp}$, and $t_{nn}$
being real numbers.  Moreover, the form of the transformation
(\ref{genbog}) has the additional consequence that $\rho_{pn}$ and
$t_{pn}$ are purely imaginary.

Now the requirements (\ref{rest2}) imply
\beqa
\rho_{pp}\,-\, \rho_{pp}^2\,-\, |\rho_{pn}|^2&=&t_{pp}^2\,+\,
 |t_{pn}|^2 , \non \\
\rho_{nn}\,-\,\rho_{nn}^2 \,-\,|\rho_{pn}|^2 &=& t_{nn}^2\,
+\,|t_{pn}|^2 , \non \\
\rho_{pn}\,(1- t_{pp}-t_{nn})&=&t_{pn}\,(t_{nn}+t_{pp}) , \non \\
\rho_{pn}\,(t_{pp}-t_{nn})&=& t_{pn}\,(\rho_{pp}-\rho_{nn}) .
\label{sagraz}
\eeqa

Let us consider the expectation value of the rising isospin operator
\beqa
<\tau_{+1}>
&=&-\frac{1}{\sqrt{2}}\sum_{jm>0}(\rho_{np}+\rho_{{\bar n}{\bar p}})
\non \\
&=&-\frac{1}{\sqrt{2}}\sum_{jm>0}(\rho_{np}+\rho_{np}^*) .
\eeqa
Thus this expectation value is always zero for transformations which
yield purely imaginary values for the non-diagonal matrix elements
$\rho_{np}$ (as (\ref{genbog}) does).

If $N_n=N_p$ and $\rho_{np}=0$, Eqs.~(\ref{sagraz}) indicate that
$t_{pp}=-t_{nn}$ (Goodman's solution).  However, although such solution
may exist, there might be also another solution with $\rho_{np}\neq 0$
and $t_{pp}=t_{nn}$.

\section{The construction of the generalized BCS basis } \label{sec:AppBB}

In the following we define the pair operators $S_v,S_0$ in the
$\lambda,\mu$ notation.
\beqa
S^{\dag}_v&=&\sum_{\lambda, \mu>0}c^{\dag}_{v\lambda\mu}
c^{\dag}_{v\lambda {\bar \mu}} , \non \\
S^{\dag}_0&=& -i\sum_{\lambda, \mu>0}
k_{\lambda\mu}(c^{\dag}_{p\lambda\mu}c^{\dag}_{n\lambda{\bar \mu} }
-c^{\dag}_{n\lambda\mu}c^{\dag}_{p\lambda{\bar \mu} }) . \label{ss}
\eeqa
The  equivalence within the $ls$ and $jj$ coupling schemes is given in
Table \ref{t1}, as well as the coupling factor $k_{\lambda \mu}$.  The
pairing operators (\ref{ss}) and the single particle operators may be
expressed in  the space of the quasi-particles using the inverse of the
transformation (\ref{genbog}).  The result for the vacuum terms is:
\beqa
<S_v>&=& \sum_{w,\lambda,\mu>0}u_{vw}v_{vw} , \non \\
<S_0>&=& \sum_{v,\lambda, \mu>0}k_{\lambda \mu}(
u_{pv}v_{nv}+u_{nv}v_{pv}), \non \\
< c^{\dag}_{v\lambda\mu} c_{v\lambda\mu} >&=& \sum_w v^2_{vw} ,
\label{eq:B2}
\eeqa
while the two quasi-particle components  appearing in the 
Hamiltonian
(\ref{qp}) yield
\beqa
(S^{\dag}_v+S_v)^{(11)}&=&-2\sum_{w,\lambda,\mu}u_{vw}v_{vw}
\alpha^{\dag}_{w\lambda\mu}\alpha_{w\lambda\mu}\non \\
&&+i\sum_{\lambda, \mu>0}(u_{vp}v_{vn}+u_{vn}v_{vp})
(\alpha^{\dag}_{p\lambda \mu}\alpha_{n\lambda \mu}-
\alpha^{\dag}_{p\lambda {\bar \mu}}\alpha_{n\lambda {\bar \mu}})+
\mbox{h.c.} , \label{svqp}
\non \\
(S^{\dag}_0+S_0)^{(11)}&=&
-2\sum_{v,\lambda,\mu}k_{\lambda\mu}(u_{pv}v_{nv}+u_{nv}v_{pv})
\alpha^{\dag}_{v\lambda\mu}\alpha_{v\lambda\mu}\non \\
&&+i\sum_{\lambda, \mu>0}k_{\lambda \mu}(u_{pp}v_{nn}+u_{nn}v_{pp}+
u_{pn}v_{np}+u_{np}v_{pn})\non \\
&&\times (\alpha^{\dag}_{p\lambda \mu}\alpha_{n\lambda \mu}-
\alpha^{\dag}_{p\lambda {\bar \mu}}\alpha_{n\lambda {\bar \mu}})+
\mbox{h.c.}  .\label{s0qp}
\eeqa
Finally, the components of the single-particle Hamiltonian are written
\beqa
\left(\sum_{v,\lambda}\epsilon_{v\lambda}\sum_\mu
c^{\dag}_{v\lambda \mu}c_{v\lambda \mu}\right)^{(11)}
&=&\sum_{v,\lambda}\epsilon_{v\lambda}\sum_{w,\mu}
(u^2_{vw}-v^2_{vw})\alpha^{\dag}_{w\lambda\mu}\alpha_{w\lambda\mu}\non \\
&&-i\sum_{v,\lambda}\epsilon_{v\lambda}
\sum_{\mu>0}(u_{vp}u_{vn}-v_{vp}v_{vn})\non \\
&&\times (\alpha^{\dag}_{p\lambda \mu}\alpha_{n\lambda \mu}-
\alpha^{\dag}_{p\lambda {\bar \mu}}\alpha_{n\lambda {\bar \mu}})+
\mbox{h.c.} . \label{sp}
\eeqa
Using (\ref{svqp}-\ref{sp}), the diagonalization of 
Eq.~(\ref{qp}) yields the two equations:
\beqa
E_{v\lambda\mu}&=&\sum_w\epsilon_{w\lambda}(u^2_{wv}-v^2_{wv})+2
\sum_w\Delta_wu_{wv}v_{wv}\non \\
&&+2\Delta_0k_{\lambda \mu}(u_{pv}v_{nv}+u_{nv}v_{pv}) , \non \\
0&=& \sum_v\epsilon_{v\lambda}(u_{vp}u_{vn}-v_{vp}v_{vn})+
\sum_v\Delta_v(u_{vp}v_{vn}+u_{vn}v_{vp})\non \\
&&+\Delta_0k_{\lambda \mu}(u_{pp}v_{nn}+u_{nn}v_{pp}+
u_{np}v_{pn}+u_{pn}v_{np}) ,
\eeqa
which are equivalent to the matrix equation (\ref{mat}).

The (positive) eigenvalues are:
\beqa
E_{v\lambda\mu}&=&\frac{1}{\sqrt{2}}\sqrt{\epsilon^2_{p\lambda}+
\epsilon^2_{n\lambda}+
\Delta^2_p+\Delta^2_n+2\Delta^2_0k^2_{\lambda \mu}+vD} , \\
D&=&\sqrt{(\epsilon^2_{p\lambda}+\Delta^2_p-\epsilon^2_{n\lambda}-
\Delta^2_n)^2+
4\Delta^2_0k^2_{\lambda\mu}\left((\epsilon_{p\lambda}-
\epsilon_{n\lambda})^2+
(\Delta_p+\Delta_n)^2
\right)} , \non
\eeqa
where we may (arbitrarily) assign the highest energy ($v$=1) to the
quasi-proton. The eigenvalues  depend on the magnetic projection $\mu$
through the factor $k_{\lambda \mu}$. Therefore, according to Table
\ref{t1}, they do so in the case of $jj$ coupling, but not for $ls$
coupling.

In this paper we are interested in the situation $\epsilon_{n\lambda}=
\epsilon_{p\lambda}$ ($=\epsilon_\lambda$) and  $N_p=N_n$
The expression for the eigenvalues simplifies to:
\beqa
E_{v\lambda\mu}=\sqrt{\epsilon_\lambda^2+\Delta_p^2+
\Delta_0^2k_{\lambda\mu}^2
+v\Delta_0k_{\lambda\mu}(\Delta_p+\Delta_n)} . \label{simple}
\eeqa

The eigenvectors of (\ref{mat})  yield the coefficients of the
quasi-particle transformation (\ref{genbog}). If the conditions leading
to (\ref{simple}) are valid,  we obtain the (unnormalized) coefficients
\beqa
u_{pv}&=&\frac{1}{2}\Delta_{0}k_{\lambda\mu}
\left(\Delta_p+\Delta_n+v\sqrt{4\Delta^2_{0}k^2_{\lambda\mu}
+(\Delta_p-\Delta_n)^2}\right) , \non \\
u_{nv}&=&\Delta^2_{0}k^2_{\lambda\mu}+\frac{1}{2}\Delta_n
\left(\Delta_n-\Delta_p+v\sqrt{4\Delta^2_{0}k^2_{\lambda\mu}
+(\Delta_p-\Delta_n)^2}\right) , \non \\
v_{pv}&=&\,\Delta_{0}k_{\lambda\mu}(E_{v\lambda\mu}-\epsilon_\lambda) ,
\non \\
v_{nv}&=&\,\frac{1}{2}(E_{v\lambda\mu}-\epsilon_\lambda)
\left(\Delta_n-\Delta_p+v\sqrt{4\Delta^2_{0}k^2_{\lambda\,u}
+(\Delta_p-\Delta_n)^2}\right) . \label{b23}
\eeqa

\section{The solutions for the case of a single $\lowercase{l}$-shell} 
\label{sec:AppCC}

\subsection{Solution with $\Delta_p=-\Delta_n$} \label{cara}

According to Eqs.~(\ref{simple}) and (\ref{b23})
\beqa
E_v&=&E=\sqrt{\epsilon^2+\Delta^2}\;;\;\;\;\;\;
\Delta=\sqrt{\Delta^2_p+\frac{\Delta^2_0}{2}} ; \non \\
u_{pv}&=&\frac{v}{2}
\sqrt{\frac{\Delta+
v\Delta_p}{E}}\left(\frac{E+\epsilon}{E-\epsilon}\right)^{\frac{1}{4}}
\;;\;\;
u_{nv}=\frac{1}{2}
\sqrt{\frac{\Delta-
v\Delta_p}{E}}\left(\frac{E+\epsilon}{E-\epsilon}\right)^{\frac{1}{4}} ;
\non\\
v_{pv}&=&\frac{1}{2}
\sqrt{\frac{\Delta+
v\Delta_p}{E}}\left(\frac{E-\epsilon}{E+\epsilon}\right)^{\frac{1}{4}}
\;;\;\;
u_{nv}=\frac{v}{2}
\sqrt{\frac{\Delta-
v\Delta_p}{E}}\left(
\frac{E-\epsilon}{E+\epsilon}\right)^{\frac{1}{4}} . \label{uvg}
\eeqa
The self-consistent equations (\ref{self}) and the number equations
(\ref{lagr}) yield
\beqa
g_1=g_0=g=\frac{2E}{2l+1}\;;\;\;\;\;\;\;\eta=
-\frac{\epsilon}{E}\label{self-} .
\eeqa
The vacuum energy (\ref{wgs}) is:
\beqa
W_{gs}=-\frac{1}{2}g(2l+1)^2(1-\eta^2) .
\eeqa

\subsection{Solution with $\Delta_p=\Delta_n$} \label{rota}

\beqa
E_v&=&\sqrt{\epsilon^2+(\Delta_p+v\Delta_0/\sqrt{2})^2}
=\Delta_p\sqrt{\zeta^2+(1+v\gamma/\sqrt{2})^2} ; \non\\
u_{pv}&=&\frac{\Delta_p+v\Delta_0/\sqrt{2}}{2\sqrt{E_v
(E_v-\epsilon)}}
\;;\;\;
u_{nv}=\frac{v\Delta_p+\Delta_0/\sqrt{2}}{2\sqrt{E_v
(E_v-\epsilon)}}
; \non \\
v_{pv}&=&\frac{1}{2}
\sqrt{\frac{E_v-\epsilon}{E_v}}
\;;\;\;\;\;\;\;\;
v_{nv}=\frac{v}{2}\sqrt{\frac{E_v-\epsilon}{E_v}} ,
\eeqa
where $\zeta,\gamma$ are given in (\ref{horn}).  The self-consistency
equations (\ref{self}) and number equations (\ref{lagr}) read
\beqa
1&=&g_1\frac{2l+1}{4E_pE_n}\left(E_p+E_n+
\frac{\gamma}{\sqrt{2}}(E_n-E_p)\right) \label{self1}\\ &=&g_0\frac{2l+1}{4E_pE_n}\left(E_p+E_n+
\frac{\sqrt{2}}{\gamma}(E_n-E_p)\right) , \label{self2}\\
\eta&=&-\frac{E_p+E_n}{2E_pE_n} \; \epsilon . \label{num}
\eeqa
A combination of the the Eqs.~(\ref{self1}) and (\ref{self2}) yields an
expression for the parameter $x$ measuring the relative strengths of
the two pairing interactions (\ref{x}).
\beqa
x=\frac{(E_n-E_p)(\frac{\gamma}{\sqrt{2}}-\frac{\sqrt{2}}{\gamma})}
{2(E_p+E_n)+(E_n-E_p)(\frac{\gamma}{\sqrt{2}}+\frac{\sqrt{2}}{\gamma})} .
\label{selfx}
\eeqa
In principle we should simultaneously solve the three equations
(\ref{self1})--(\ref{num}) in order to obtain $\Delta_p, \Delta_0,
\epsilon$ as functions of $g_1,g_0,\eta$. However it is easy to verify
that (\ref{selfx}) and (\ref{num}) depend only on the two  variables
$\gamma$ and $\zeta$. Therefore it is simpler to solve first the system
of two equations and subsequently obtain the values of
$\Delta_p,\Delta_0$ through an independent combination of Eqs.
(\ref{self1}) and (\ref{self2}). This last step yields
\beqa
\Delta_p&=&\frac{g(2l+1)}{4E_pE_n} \times \non \\
&&\sqrt{(1-x^2)
\left(E_p+E_n+\frac{\gamma}{\sqrt{2}}(E_n-E_p)\right)
\left(E_p+E_n+\frac{\sqrt{2}}{\gamma}(E_n-E_p)\right)} , \non \\
\Delta_0&=& \Delta_p\, \gamma . \label{cam}
\eeqa
These gaps may be used to obtain the  vacuum energy $W_{gs}$ (cf.
(\ref{wgs})).

\subsubsection{Symmetries and domain of existence} 
\label{sucia}

The system of Eqs.~(\ref{num}) and (\ref{selfx}) remains
invariant under the two following transformations
\beqa
\frac{\gamma}{\sqrt{2}}&\rightarrow& \frac{\sqrt{2}}{\gamma}\;;\;\;\;\;
\zeta \rightarrow \frac{\zeta\sqrt{2}}{\gamma}\;;\;\;\;\; x\rightarrow -x\;;\;\;\;\; \eta\rightarrow \eta ,
\label{inv1} \\
 \gamma&\rightarrow &\gamma\;;\;\;\;\; \zeta\rightarrow -\zeta\;;\;\;\;\; 
x\rightarrow x\;;\;\;\;\; \eta\rightarrow -\eta . \label{inv2}
\eeqa

As a consequence of the invariance under the transformation (\ref{inv1}) 
\beqa
\Delta_p\rightarrow \frac{\Delta_0}{\sqrt{2}}\;;\;\;\;\;\;
\Delta_0 \rightarrow \sqrt{2} \Delta_p . \label{sire}
\eeqa
Thus this invariance implies
\beqa 
W_{gs}(x,\eta) \rightarrow W_{gs}(-x,\eta) . \label{zono}
\eeqa

Since  $\Delta_p$ and $ \Delta_0$ remain invariant under the 
transformation (\ref{inv2}),
\beqa
W_{gs}(x,\eta) \rightarrow W_{gs}(x,-\eta) .
\eeqa

In order to find the region of allowed solutions in the $(x,\eta)$
plane we discuss the two limits $\zeta\rightarrow \infty$ and
$\zeta\rightarrow 0$.

For the first limit, Eq.~(\ref{num}) requires that $\gamma=\beta_\infty
\zeta$ (with $\beta_\infty$=constant). We obtain
\beqa
\lim_{\zeta\rightarrow \infty} x= -\frac{\beta^2_\infty}{2+\beta^2_\infty}\;;\;\;\;\;\;
\lim_{\zeta\rightarrow \infty} 
\eta=-\frac{1}{\sqrt{1+\beta^2_\infty}} . \label{infy}
\eeqa
The consistency between these two equations yields the curve $\eta_1(x_1)$ (\ref{panda})
limiting the region of allowed solutions.  Let us calculate now the
gaps and the vacuum energy on this curve. From Eqs.~(\ref{cam}) we
obtain
\beqa
\lim_{\zeta\rightarrow \infty} \Delta_p&=&0 , \non \\
\lim_{\zeta\rightarrow \infty} \Delta_0&=&(2l+1)
\,\frac{g}{\sqrt{2}}\,
\sqrt{1-x_1^2}\,\frac{\beta_\infty}{1+\beta_\infty^2}
\;;\;\;\;\;(-1\leq x_1\leq 0) \non\\
&=& (2l+1)\,\frac{g}{\sqrt{2}}\,(1+x_1)\,\sqrt{1-\eta^2_1} ,
\label{cord}
\eeqa
which implies
\beqa
W_{gs}
&=&-(2l+1)^2\,\frac{g(1+x_1)}{2}\,(1-\eta^2_1)\non \\
&=& -(2l+1)^2\, \frac{g_0(x_1)}{2}\,(1-\eta^2_1)  . \label{cebra}
\eeqa
In order to obtain the energy for positive values of $x_1$ we apply
again the transformation (\ref{inv1}).  According to (\ref{zono}),
\beqa
W_{gs}&=&-(2l+1)^2\,\frac{g(1-x_1)}{2}\,(1-\eta^2_1)\non \\
&=&-(2l+1)^2\, \frac{g_1(x_1)}{2}\,(1-\eta^2_1) . \label{agui}
\eeqa

Let us consider now the limit $\zeta\rightarrow 0$. Similarly to the
previous limiting procedure, we  require  $\gamma\rightarrow \sqrt{2}
\beta_0 \zeta$, with $\beta_0$ constant. The limiting procedure yields
\beqa
\lim_{\zeta\rightarrow 0} x =\frac{\beta_0}{\sqrt{1+\beta^2_0}}\;;\;\;\;\;
\lim_{\zeta\rightarrow 0} \eta =  -\frac{1}{2\sqrt{1+\beta^2_0}} .
\label{cero}
\eeqa
The last equation implies that $\beta_0$ is a real number in the
interval $-\frac{1}{2}\leq \eta \leq \frac{1}{2}$. Within this interval
we obtain the curve (\ref{oso}), which is complementary to
(\ref{panda}) and (\ref{pandap}).

\subsection{The two trivial solutions}\label{fea}

The results are summarized in Table \ref{t2}.

\section{The solutions for the case of a single $\lowercase{j}$-shell} 
\label{sec:AppDD}

The pairing operators used in Section \ref{singlel} and in Appendix
\ref{sec:AppCC} are given explicitly in the second column of Table
\ref{t1}. They may also be written in the $jj$ coupling scheme,
\beqa
(S^{01}_{0M_t})^{\dag}&=&
-\sqrt{\frac{2l+1}{2}}[c^{\dag}_lc^{\dag}_l]^{L=0,J=0,T=1}_{0,0,M_t}
=-\sum_{j=j^+,j^-}\frac{\sqrt{2j+1}}{2}
[c^{\dag}_jc^{\dag}_j]^{J=0,T=1}_{0,M_t} ,
\non \\
(S^{10}_{M0})^{\dag}&=&i\sqrt{\frac{2l+1}{2}}
[c^{\dag}_lc^{\dag}_l]^{L=0,J=1,T=0}_{0,M,0}
\non \\
&=&i\sqrt{\frac{(2j^++1)(j^++1)}{6j^+}}
[c^{\dag}_{pj^+}c^{\dag}_{nj^+}]^{J=1}_{M}\non \\
&&-\,i\sqrt{\frac{(2j^-+1)j^-}{6(j^-+1)}}
[c^{\dag}_{pj^-}c^{\dag}_{nj^-}]^{J=1}_{M}
\non \\
&&-\,i\sqrt{\frac{(2j^++1)(2j^-+1)}{6j^+}}
\left([c^{\dag}_{pj^+}c^{\dag}_{nj^-}]^{J=1}_{M}-
[c^{\dag}_{nj^+}c^{\dag}_{pj^-}]^{J=1}_{M}\right) ,
\eeqa
where $j^{\pm}=l\pm \frac{1}{2}$. In this Appendix we consider the case
of a single $j$-shell which may be derived from the previous $l$-case
by including a large spin-orbit term in the Hamiltonian. In such case
we may keep only the  contribution to $(S^{JT}_{MM_t})^{\dag}$
corresponding to the valence shell. We multiply  this contribution to
$(S^{10}_{M0})^{\dag}$ by a  factor (chosen for convenience) that
yields the expression given in the third column of Table \ref{t1}.

\subsection{Solution with $\Delta_p=-\Delta_n$} \label{caro}

According to Eqs.~(\ref{simple}) and (\ref{b23})
\beqa
E_v&=&E=\sqrt{\epsilon^2+\Delta^2}\;;\;\;\;\;\;\Delta=\sqrt{\Delta^2_p
+\frac{2m^2\Delta^2_0}{j(j+1)}} ; \non
 \\
u_{pv}&=&\frac{v}{2}
\sqrt{\frac{\Delta+
v\Delta_p}{E}}\left(\frac{E+\epsilon}{E-\epsilon}\right)^{\frac{1}{4}}
\;;\;\;
u_{nv}=\frac{1}{2}
\sqrt{\frac{\Delta-
v\Delta_p}{E}}\left(\frac{E+\epsilon}{E-\epsilon}\right)^{\frac{1}{4}}
; \non\\
v_{pv}&=&\frac{1}{2}
\sqrt{\frac{\Delta+
v\Delta_p}{E}}\left(\frac{E-\epsilon}{E+\epsilon}\right)^{\frac{1}{4}}
\;;\;\;
u_{nv}=\frac{v}{2}
\sqrt{\frac{\Delta-
v\Delta_p}{E}}\left(
\frac{E-\epsilon}{E+\epsilon}\right)^{\frac{1}{4}} . \label{uvh}
\eeqa
The self-consistent conditions and number equation are:
\beqa
1&=&\frac{1}{4}g_1s_0=\frac{1}{2}g_0s_2\;;\;\;\;\;\;
\eta=-\frac{\epsilon s_0}{2j+1} , \non \\
s_i&=&\sum_{m>0}\left(m\sqrt{\frac{2}{j(j+1)}}\right)^i
\left(\frac{1}{E_p}+(-1)^i\frac{1}{E_n}\right) .
\label{hiena}
\eeqa

In the limit $\Delta_0\rightarrow 0$,
the self consistent conditions yield the
limiting value $x\geq -1/7$.

\subsection{Solution with $\Delta_p=\Delta_n$}  \label{roto}
\beqa
E_v&=&\sqrt{\epsilon^2+
\left(\Delta_p+v\Delta_0m\sqrt{\frac{2}{j(j+1)}}\right)^2} ;
\label{gato} \\
u_{pv}&=&\frac{\Delta_p+v\Delta_0m\sqrt{\frac{2}{j(j+1)}}}{2\sqrt{E_v
(E_v-\epsilon)}}
\;;\;\;
u_{nv}=\frac{v\Delta_p+\Delta_0m\sqrt{\frac{2}{j(j+1)}}}{2\sqrt{E_v
(E_v-\epsilon)}} ;
\non \\
v_{pv}&=&\frac{1}{2}
\sqrt{\frac{E_v-\epsilon}{E_v}}
\;;\;\;\;\;\;\;\;
v_{nv}=\frac{v}{2}\sqrt{\frac{E_v-\epsilon}{E_v}} . \non
\eeqa
The self-consistency equations (\ref{self}) and number equations
(\ref{lagr}) read
\beqa
1=\frac{1}{4}g_1(s_0+\gamma s_1)=\frac{1}{2}
g_0(\frac{s_1}{\gamma}+ s_2) \;;\;\;\;\;\;
\eta =-\frac{\epsilon s_0}{2j+1} , \label{sc} 
\eeqa
where the $s_i$ are given in Eq.~(\ref{hiena}).

\subsubsection{The analytical treatment of the solution 
in the middle of the shell ($\eta=\epsilon=0$)}
\label{roto2}

According to Eq.~(\ref{gato}), the quasi-particle energies
are ($m>0$):
\beqa
E_p=\Delta_p\,\left(1+\gamma \mj\right)\;;\;\;\;\;\;
E_n=\Delta _p\,\left|1-\gamma \mj\right| , \label{jil}
\eeqa
which yields the results of Table \ref{lombriz}. With these results we
may evaluate the expressions appearing in the self-consistency
conditions (\ref{sc})
\beqa
s_0+\gamma s_1&=& \frac{2}{\Delta_p}\,\sum_{m>0}
\,y_m , \non \\
s_1+\gamma s_2&=& \frac{2}{\Delta_p}\,\sum_{m>0}
\,\mj\,(1-y_m) ,
\eeqa
where the values of $y_m$ are also given in Table \ref{lombriz}.
Therefore, the relative strength $x$ has the value:
\beqa
x=\frac{
- (j+\frac{1}{2})^2\,+\,
\sum_{m>0}\,y_m\,
\left(\gamma\sqrt{\frac{j(j+1)}{2}}+
2 m \right)}
{ (j+\frac{1}{2})^2\,+\,
\sum_{m>0}\,y_m\,
\left(\gamma\sqrt{\frac{j(j+1)}{2}}-
2 m \right)} , \label{pardo}
\eeqa
which clearly displays the  discontinuities to be expected due to the
form of $E_n$ at
\beqa
1=\gamma \mj . \label{disc}
\eeqa
In the interval $k$  $(k=m+\frac{1}{2})$ defined as
\beqa
\frac{\sqrt{2j(j+1)}}{2k-1}\,>\,\gamma\,>\,
\frac{\sqrt{2j(j+1)}}{2k+1} \label{inter}
\eeqa
Eq.~(\ref{pardo}) is equivalent to
\beqa
x=\frac{\gamma k\sqrt{j(j+1)/2}- 
\left((j+\frac{1}{2})^2-k^2\right)}
{\gamma k\sqrt{j(j+1)/2}+ 
\left((j+\frac{1}{2})^2-k^2\right)} ,
\eeqa
which may be inverted so as to yield the ratio $\gamma$ as a function
of the relative coupling strength~$x$
\beqa
\gamma=\left(\frac{1+x}{1-x}\right)\, 
\frac{(j+\frac{1}{2})^2-k^2}{k\sqrt{j(j+1)/2}} . \label{card}
\eeqa
The value of $\gamma$ is represented in Fig.~\ref{f5} as a function of
$x$. The discontinuities are apparent. Note also that for $x<0$ there
are several values of $\gamma$ which are compatible with a single value
of $x$. The opposite situation appears for $x>0$, in which case there
may not be  values of $\gamma$ associated with some values of $x$.
 
Within the same interval we may express the vacuum energy $W_{gs}$ as a
function of $x$
\beqa
W_{gs}=-\frac{g}{2}
\left((1-x)k^2-(1+x)\frac{(k^2-(j+\frac{1}{2})^2)^2}
{j(j+1)}\right) , \label{gorr}
\eeqa
which displays the linearity in $x$ within each interval. This
behavior was already found in Fig.~\ref{f4} and it is to
be expected for degenerate shells.

The initial and final values $x_{\pm}$, for each interval $k$, may also
be given as a function of $k$
\beqa
x_{\pm}=\frac{kj(j+1)-(2k\mp 1)\left((j+\frac{1}{2})^2-k^2\right)}
{kj(j+1)+(2k\mp 1)\left((j+\frac{1}{2})^2-k^2\right)} . \label{pal}
\eeqa

Let us study now the behavior: i) before the first interval starts at
$m=\frac{1}{2}$ and ii) after the last one ends at $m=j$. Note that
$\gamma$ not only decreases from one interval to the next one
(\ref{disc}) but also within each interval (\ref{inter}). The
corresponding values are given in Table \ref{abeja}. The value $x=-1$
($+1$) becomes associated with  a predominant isoscalar (isovector)
pairing correlation.  This is consistent with the fact that the
solution with $\Delta_p=\Delta_n$ merges, at the extremes of the
$x$-interval, with the most unfavored trivial solution 
having the same
value of $x$.

\subsection{The trivial solutions} \label{tribu}

The solution with $\Delta_0=0$ is the familiar one 
with
\beqa
E&=&\sqrt{\epsilon^2+\Delta^2_p}\;;\;\;\;\;\;
\epsilon=-\eta\, E ;
\non \\
\Delta_p&=& -\frac{g_1}{2}\,(j+\frac{1}{2})\,\sqrt{1-\eta^2} ; \non \\
W_{gs}&=&-\frac{g_1}{2}\,(j+\frac{1}{2})^2\,(1-\eta^2) .
\eeqa

The case with $\Delta_p=0$ requires to solve the equation
\beqa
\eta=-\frac{1}{j+\frac{1}{2}}\,\frac{\epsilon}{\Delta_0}\,
\sum_{m>0}
\frac{1}{\sqrt{(\epsilon / \Delta_0)^2
+2m^2 / j(j+1)}} . \label{chac}
\eeqa
With the value of the ratio $\epsilon/\Delta_0$ thus determined, one
obtains the gap and the vacuum energy.
\beqa
\Delta_0=\frac{2g_0}{j(j+1)}\sum_{m>0}
\frac{m^2}{\sqrt{(\epsilon / \Delta_0)^2
+2m^2 / j(j+1)}}\;;\;\;\;\;\;W_{g.s}=-\frac{\Delta_0^2}{g_0} .
\eeqa

The solution becomes simplified for the case
$\eta=\epsilon=0$, namely,
\beqa
\Delta_0(\epsilon=0)&=&\frac{g_0}{2}\,(j+\frac{1}{2})^2
\sqrt{\frac{2}{j+1}} , \non \\
W_{gs}(\epsilon=0)&=& -\frac{g_0}{2}\, \frac{(j+\frac{1}{2})^4}{
j(j+1)} .
\eeqa

\vspace*{2cm}

\newpage

\mediumtext
\begin{table}[p]
\begin{tabular}{c|c|c}
$\lambda$ &$l$       &$j$\\
$\mu$ &$m,\sigma=\pm \frac{1}{2}$ &$m$\\
$\mu>0$ & $-l\leq m \leq l$, $\sigma=\frac{1}{2}$ & $m>0$\\
$k_{\lambda\mu}$& $\frac{1}{\sqrt{2}}$ 
&$m\sqrt{\frac{2}{j(j+1)}}$\\
\hline
$S^{\dag}_v$ &$-\sum_l\sqrt{\frac{2l+1}{2}}
[c^{\dag}_lc^{\dag}_l]^{L=0,J=0,T=1}_{T_0=v}$
&$-\sum_j\frac{\sqrt{2j+1}}{2}
[c^{\dag}_jc^{\dag}_j]^{J=0,T=1}_{T_0=0}$\\
$S^{\dag}_0$&$i\sum_l\sqrt{\frac{2l+1}{2}}
[c^{\dag}_lc^{\dag}_l]^{L=0,J=1,T=0}_{J_0=0}$&
$i\sum_j\sqrt{\frac{2j+1}{3}}
[c^{\dag}_jc^{\dag}_j]^{J=1,T=0}_{J_0=0}$\\
\hline
$\eta $&$ \frac{N_p}{2l+1}-1$ & $\frac{N_p}{j+\frac{1}{2}}-1$ 
\end{tabular}
\vspace{.5cm}
\caption{The equivalence of the notation $\lambda,\mu$ in the $ls$ and
$jj$ coupling schemes,  the definition of the pairing operators
$S_{v,0}^{\dag}$ and of the occupation number $\eta$ ($-1 \leq \eta \leq
1$).}
\label{t1}
\end{table}

\narrowtext
\begin{table}[p] 
\begin{tabular}{c|c|c}
$\Delta_p$ & $\frac{1}{2}g_1(2l+1)\sqrt{1-\eta^2}$& 0 \\
$\Delta_0 $& 0 &$ \frac{1}{\sqrt{2}}g_0(2l+1)\sqrt{1-\eta^2}$  \\
$\epsilon $& $-\eta E $& $-\eta E$ \\
$E=E_p=E_n$ & $\frac{1}{2} g_1(2l+1)$  & $\frac{1}{2} g_0(2l+1)$ \\
$W_{gs}$  & $-\frac{1}{2}g_1(2l+1)^2(1-\eta^2)$&$ 
-\frac{1}{2}g_0(2l+1)^2(1-\eta^2)$
\end{tabular}
\vspace{.5cm}
\caption{The solutions with $\Delta_0=0$ (second column) and
$\Delta_p=0$ (third column) for a single $l$-shell.}
\label{t2}
\end{table}

\mediumtext
\begin{table}[p]
\begin{tabular}{c|c|c}
 & $1>\gamma \mj$ & $1< \gamma \mj$ \\
\hline
$\frac{1}{E_p}+\frac{1}{E_n}$& 
$2/\Delta_p(1-\frac{2m^2\gamma^2}{j(j+1)})$& 
$-2\gamma \mj /\Delta_p(1-\frac{2m^2\gamma^2}{j(j+1)})$\\
$\frac{1}{E_p}-\frac{1}{E_n}$&
$-2\gamma \mj /\Delta_p(1-\frac{2m^2\gamma^2}{j(j+1)})$&
$2/\Delta_p(1-\frac{2m^2\gamma^2}{j(j+1)})$\\
\hline
$y_m$ &1 &0
\end{tabular}
\vspace*{.5cm}
\caption{The values of $\frac{1}{E_p}\pm\frac{1}{E_n}$ and of $y_m$ as
functions of $\gamma$}
\label{lombriz}
\end{table}

\narrowtext
\begin{table}[p]
\begin{tabular}{c|c|c|c|r}
$\gamma$& $y_m $ & $\sum_{m>0} y_m$ & $\sum_{m>0}my_m$& $x$\\
\hline
$\sqrt{2j(j+1)}<\gamma<\infty$ & 0 & 0 & 0 & -1\\
$0<\gamma<\sqrt{\frac{j+1}{2j}}$ & 1 & $j+\frac{1}{2}$ &
$ \frac{1}{2}(j+\frac{1}{2})^2 $& 1
\end{tabular}
\vspace*{.5cm}
\caption{The behavior of the solution for very large and very small
values of $\gamma$}
\label{abeja}
\end{table}

\end{document}